\documentclass[twocolumn,showpacs,preprintnumbers,amsmath,amssymb]{revtex4}
\usepackage{graphicx}
\usepackage{dcolumn}
\usepackage{bm}

\begin{document}

\title{Bands, spin fluctuations and traces of Fermi surfaces in angle-resolved photoemission intensities for
high-$T_C$ cuprates.}

\author{T. Jarlborg}

\affiliation{
DPMC, University of Geneva, 24 Quai Ernest-Ansermet, CH-1211 Geneva 4,
Switzerland
}


\begin{abstract}

The band structures of pure and hole doped La$_2$CuO$_4$ with
antiferromagnetic (AFM) spin-fluctuations are calculated and
compared to spectral weights of ARPES.
It is shown that observations of coexisting Fermi surface (FS) arcs and
closed FS pockets are consistent with modulated AFM spin fluctuations 
of varying wave lengths. 
Large variations of strong spin fluctuations make the outer part of the FS break
diffuse at low doping. This part of the FS is suppressed at high doping when 
spin fluctuations are weak.
The resulting superimposed spectral weight has features both from FS arcs and closed pockets.
A connection between results of ARPES, neutron scattering, and band results for
the modulated AFM spin wave state, suggests
that spin-phonon coupling is an important mechanism for the
properties of the cuprates.

\end{abstract}


\maketitle

\section{Introduction.}

The spectral weights of angle-resolved photoemission spectroscopy (ARPES) in moderately
hole doped high-$T_C$ cuprates indicate that the Fermi surface (FS) is
incomplete, and forms a "FS-arc" that becomes shorter as the temperature
is lowered \cite{dama,nor,kani,meng,hwang}. In Bi$_2$Sr$_2$CaCu$_2$O$_{(8-\delta)}$ 
the FS is not seen near the the $X$-point $(\pi,0)$, which is taken as a signature
of an energy gap \cite{mars}, and a similar
behavior in Ca$_{(2-x)}$Na$_x$CuO$_2$Cl$_2$
has been interpreted as an effect of charge ordering \cite{shen}.
An incomplete FS would be incompatible with what
is known for normal Fermi liquids, since FSs are expected to form closed
electron or hole pockets. On the other hand, the ARPES data show sometimes
double arcs (like bananas), which are consistent with the existence of closed
orbits seen from quantum
oscillations in high magnetic field \cite{yell}. Oscillations in doped YBa$_2$Cu$_3$O$_7$
(YBCO) correspond to a FS area that is much smaller ($\sim 3$ percent) \cite{doir} than the area 
of the large M-centered FS cylinder found in conventional band structure of many cuprates.
Unexpectedly, observations of Hall resistance found evidence of electron character for the pocket \cite{leBoeuf},
although it would seem more natural with a hole pocket for that part of the band structure
in hole-doped La$_{(2-x)}$Sr$_x$CuO$_4$ (LSCO) and YBCO. 

Another source of important information about the cuprates has been provided by
inelastic neutron scattering, where "hourglass"-shaped ($q,\omega$)-dependences of 
 spin excitations are
seen \cite{vig,tran1,mat,hin}. The narrow part with the shortest q-vectors has  
energies of the same order as the highest
phonon energies. Larger q-vectors are found at low energy, and even more
so at high energy before the disappearance of spin excitations at
even higher $\hbar\omega$.
These facts concerning FS features and q-dependence of spin fluctuations seem disconnected,
but there might exist a link, which even could be revealing for the mechanisms of superconductivity. 
While it can be understood that antiferromagnetic (AFM) order can break up
a large FS into smaller orbits
 \cite{harr},
it is not evident that ARPES should only detect remnants of such orbits at low doping \cite{meng,king}.
It has been argued that smearing due to finite correlation length, $\ell$, of the order 10-20 \AA (related to
lifetime $\tau$ of spin fluctuations) leads
to effects consistent with observations \cite{harr}. In the present work it is shown that also
multiple wave lengths of striped spin fluctuations, like in the observations from
neutron scattering, lead to an asymmetric FS-smearing.
The real-space picture of these fluctuations with vector $\vec{q}$ is that stripe-like potential modulations
exist, where $L \approx \pi/q$ is the thickness or wave length of a stripe \cite{acmp}. The modulations
can be static, as suggested for $T_C$-suppressed 1/8 doped LSCO \cite{mood}, or more likely dynamic in general cases.
Thus, it is suggested that the observation of asymmetric FS-orbits in ARPES is directly connected to multiple
q-vectors of spin fluctuations seen in neutron scattering. This provides a clue to the 
mechanisms of superconductivity, since spin fluctuations with multiple q-excitations
are some of the consequences of spin-phonon coupling, SPC. This coupling makes spin waves
stronger if they co-exist with phonons of the correct phase and q-vector \cite{tj9,egami}. 
The light O-modes are more efficient to enhance the spin waves than the phonons involving heavier
elements. Model calculations where the SPC varies with $q$ lead to a spin excitation
spectrum quite like what is observed. The effects on the FS from fluctuations with a range
of q-vectors as in the SPC model \cite{tj9} 
will be tested here. 

\section{Results and discussion.}

\begin{figure}
\includegraphics[height=6.0cm,width=8.0cm]{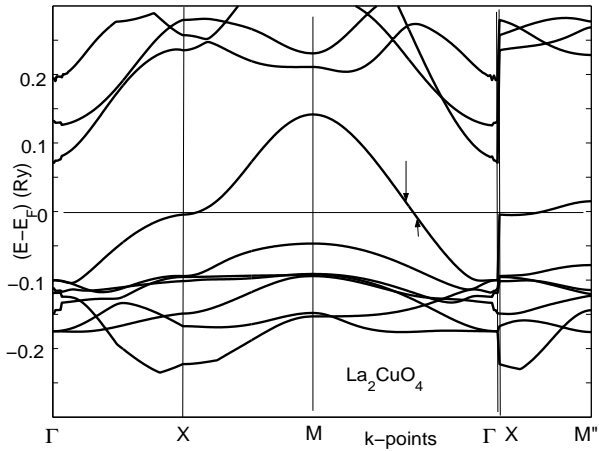}
\caption{The bandstructure (nonmagnetic) of 
La$_2$CuO$_4$ for $k_z$=0. AFM order will fold the Brillouin
zone, so that the $X-M$ bands are mirrored back on $X-\Gamma$.
The bands from $M$ to $\Gamma$ will be folded at the new zone limit
$M"$ ($\pi/2a_0,\pi/2a_0$), indicated by the downward arrow. When AFM is modulated into stripes
the folding moves closer toward $\Gamma$ (indicated by the upward arrow)
and the zone limit at $X$ is reduced correspondingly.
}
\label{fig1}
\end{figure}

\begin{figure}
\includegraphics[height=6.0cm,width=8.0cm]{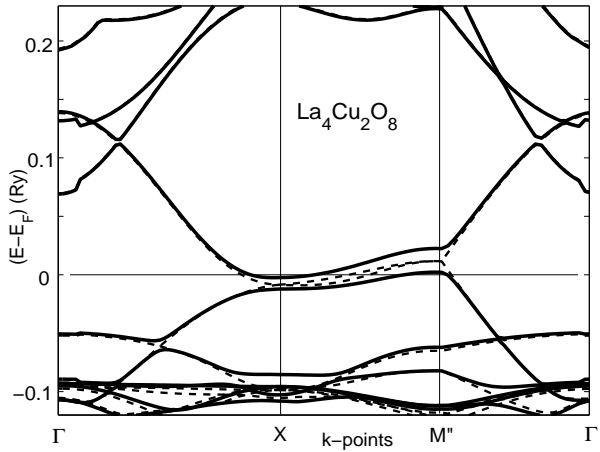}
\caption{The bandstructure of nonmagnetic (broken lines)
and AFM (full lines) La$_4$Cu$_2$O$_8$ in the AFM brillouin
zone for $k_z$=0. The AFM configuration is obtained by
having a staggered magnetic field on Cu ($\pm$7 mRy). The moments
are $\pm$0.19 $\mu_B$/Cu.
}
\label{fig2}
\end{figure}

Different experiments are made for different members of the cuprate family.
Some are difficult to cleave into clean surfaces apt for ARPES,
and photoemission is mostly made on Bi-based cuprates \cite{meng,nor,dama}. 
Neutron data are taken from large volume samples usually on La based Cu-oxides \cite{vig,tran1,mat},
and quantum oscillations were detected in crystals of YBaCuO materials \cite{yell}. However, this
diversity of systems for the data collection should not hamper an unified
understanding of the cuprates, because their structures and electronic properties are
very similar. For instance,
almost identical CuO planes are always present, on which AFM moments can appear on Cu, and
 calculated band structures
for different cuprates show the presence of at least one large cylindrical
FS ("barrel"). This FS is seen in ARPES, but the intensity 
is weak for states with k-vectors parallel to
the CuO-bond direction, which indicates a gap in this part of the Brillouin zone (BZ). 
Ultimately, the arc becomes just a point on the diagonal direction at the lowest T \cite{nor}.
The calculated density-of-state (DOS) at the Fermi energy ($E_F$) is dominated by
the Cu-d electrons from the CuO planes.  
The present study is based on the band structure of La$_2$CuO$_4$ (LCO). The calculations are made
using the Linear Muffin-Tin Orbital (LMTO) method and the
local spin density approximation.
The details of these calculations have been published earlier \cite{tj1,tj6,tj7}.
 The nonmagnetic (NM) band structure is shown in Fig. \ref{fig1}.
A single barrel-like FS is centered at $M$ (at $\pi/a_0,\pi/a_0$, where $a_0$ is the in-plane
lattice constant). AFM order on the Cu lattice will double the 
unit cell, and the bands are split by exchange and folded into a half as big AFM BZ. 
This can be followed by comparing the original bands in Fig. \ref{fig1} with the bands calculated for
the AFM cell shown in Fig. \ref{fig2} within the AFM BZ. It can be noted that the exchange
splitting, $\xi$, is largest near $E_F$ (on the $X-M"$ line) while it is lower below $E_F$.
The original barrel is fragmented into two pieces by AFM.
One banana-like feature at $M"$ (which might look like an arc if it is narrow), and another
rounded piece at $X$. The latter is delicate by two reasons. First, the band is very flat (a heavy effective mass)
at $X$, which makes the FS sensitive to disorder and fluctuations. Secondly, it 
disappears at high hole doping when $E_F$ is lower. It does not have a dispersive band
as the other piece has from $M"$ to $\Gamma$, see Fig. \ref{fig2}. A dispersive band has a sharp
FS break which is easy to detect in photoemission. But at the end points of the
banana, between $M"$ and $X$, the band is flat and the FS break is diffuse. 
An insulating gap will be opened at $E_F$ for increased exchange splitting,
when both the hole pocket at $M"$ and the electron pocket at $X$ disappear. However, the $X$ pocket 
is the first to disappear because of hole doping and a low $E_F$.
Such cases are considered in the present model calculations.

\begin{figure}
\includegraphics[height=5.5cm,width=7.0cm]{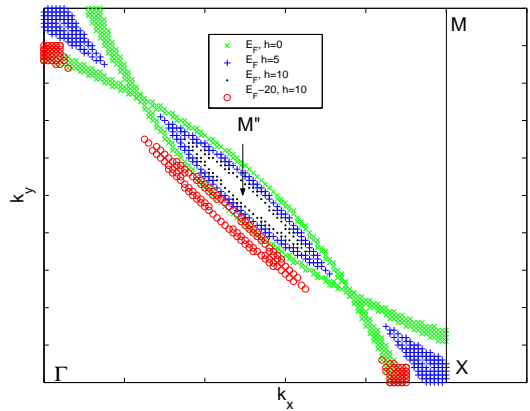}
\caption{(Color online) The evolution of the AFM FS for increased exchange splitting
(h=0,5,10 mRy) are shown by green "x", blue "+" and black ".", respectively. 
The original "unfolded" FS for h=0 has
only one circular FS centered around the M-point ($\pi/a_0,\pi/a_0$).
Two disconnected FS are 
formed when h increases. One at $X$, which is diffuse and disappears at high doping, 
and one at $M"$ with clear FS breaks toward $\Gamma$ and $M$. This piece of FS
is displaced toward $\Gamma$ when AFM is $q$-modulated 
into stripes  (red "o"). 
}
\label{fig3}
\end{figure}

The AFM order on Cu neighbors separated by a distance $a_0$ (wave vector $\vec{Q}=\pi/a_0$) can be modeled
by a nearly free electron (NFE) potential perturbation $V(x)= \xi_0 exp(i \vec{x}\cdot\vec{Q})$,
where $\xi_0$ is the maximum of the exchange splitting \cite{tj6}.
The potential for a stripe modulation with period 
$N \cdot a_0$ (wave vector $q= \vec{Q}/N$) is obtained by multiplying the potential
by $exp(-i \vec{x}\cdot\vec{q})$. Totally this gives 

\begin{equation}
 V(x) = \xi_0 e^{i \vec{x}\cdot(\vec{Q}-\vec{q})}
\label{eq1}
\end{equation}
and the zone boundary will regress from $\pi/a_0$ to 
$\pi (1-1/N)/a_0$.
The gap in a NFE model based on this potential will follow the zone boundary, and the gap
moves down in energy when $\vec{q}$ increases \cite{tj6,acmp}.
Ab-initio band calculations confirm that the gap moves toward lower energy when
the modulation is made within longer and longer unit cells, and the pseudogap at $E_F$
is a result of the striped modulations \cite{tj1,tj7}.
 By folding
the bands of Fig. \ref{fig1} it is possible to reconstruct the AFM bands, like the ones in Fig. \ref{fig2}.
Plots of the FS are made by tracking k-points with $|E(k)-E_F| < 2 mRy$ within a
BZ of 100 by 100 k-points in the $k_z =0$ plane.
The reconstructed
FS orbits are displayed in Fig. \ref{fig3}, where one case with modulated AFM is shown
by the displaced orbit. Effects of finite $\tau$ or spread
of modulated q-vectors are not yet included, and therefore, both sides of the FS are equally sharp. 

The effect of finite correlation length for quasistatic AFM modulations and
$\tau$ has been shown to make an asymmetric smearing of ARPES intensities \cite{harr}.
The result on the FS is similar in the present work, but the origin is here 
directly connected to the observed spin wave excitations.
A distribution of q-vectors makes the smearing
different on the inner (toward $\Gamma$) 
and outer FS break. 
Different q-vectors can be characteristic for different domains in the material, where local variations of
the doping make the spin fluctuations weak or strong. Such domains might very well be larger
than the typical correlation length, and here it is the q-value and not the correlation length 
that determines the broadening. 
Emitted electrons in 
ARPES come from
a wide area compared to the size of the domains, and an average of the different $\vec{q}$ from the
domains will be detected. Life-time broadening will
not contribute for static domains. Or more likely, if
fluctuations of different $\vec{q}$ are continuously mixed in space and time, there will be additional
smearing from $\tau$, since the life-time of the electronic states is shortened. The present
model will not distinguish between these two possibilities since effects of multiple $\vec{q}$
exist static and dynamic spin waves. Other effects such as diffusion at impurities and phonons will
shorten $\tau$ and make all parts of the FS diffuse. The effects of multiple $q$ are very
different on the inner and outer FS branches, such as seen experimentally. Thus, together
with the expected doping dependence, it is possible to separate standard FS broadening
effects from q-wave effects.

Neutron scattering has shown that the q-vectors of spin excitations vary from 0.05 ($\pi/a_0$ units) at intermediate
energy to 0.15 at low energy or even 0.2 for energies clearly above phonon energies \cite{vig}.
However, the intensity decreases rapidly above 60 meV, so spin fluctuations
are mainly found for $\hbar\omega$ in the range of phonon energies. Furthermore,
all q-values decrease at lower doping \cite{tran1}. Such behaviors agree with
models of spin-phonon coupling, where
strong spin fluctuations coupled to oxygen phonon modes lead to small q-vectors and vice-versa
for interaction with low frequency modes \cite{tj9}. Within the range $\hbar \omega \sim$
10-60 meV and  $\vec{q} \sim$  0.05-0.15, the amplitudes of 
$\xi_0$ are 20-30 mRy in the model for 
spin-phonon coupling. Hence, the q-vectors can vary a factor of 3, when the exchange splitting
varies $\pm$20 percent, and the effect of smearing
coming from exchange shifts of the band is rather small in comparison to smearing 
from q-variations.

\begin{figure}
\includegraphics[height=6.0cm,width=7.0cm]{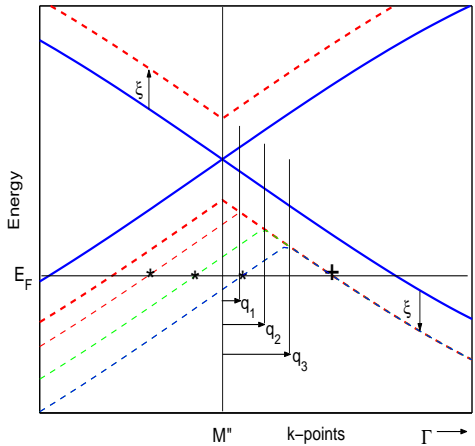}
\caption{(Color online) A schematic close-up of the bands near the M"-point.
The non-spinpolarized band is reflected at the zone boundary at M" 
to create the band structure for weak AFM (full lines). An exchange energy $\xi$ will
split the bands (heavy broken lines) and displace the FS-breaks closer to M".
 The bands from 3 striped AFM waves, modulated with 
3 different q-vectors ($q_1,q_2$ and $q_3$), will be reflected by the 3 new zone limits indicated by the
short vertical lines, as described in the text. 
These bands (thin broken lines) make FS crossings at different positions (*) on the
part outside the zone limits, while they are all superimposed on one FS crossing (+) on inside
part, i.e. the part closest to $\Gamma$.
}
\label{fig4}
\end{figure}

The inner part of the FS (the one closest
to $\Gamma$) can only be smeared if there are variations of $\xi$. But as 
the variation of $\xi$ within the band is rather small, and $\xi$ becomes
smaller further away from $E_F$, it leads only to small 
broadening effects of the inner band. On the other hand,
the FS crossing of the outer part can be shifted much more depending on
how the band is reflected in a multitude of zone boundaries.  
This is sketched in Fig. \ref{fig4}, where the heavy full (nonmagnetic) and broken  
(exchange splitted) lines show the band dispersion
in the AFM zone near $M"$ on a line of k-points going from the zone center in the 2nd BZ
(i.e. $M$ in the normal BZ) toward
$\Gamma$. 
 When disorder in form of three modulated stripes with new reduced zone boundaries
are introduced, and $\xi$ is constant, there will be three FS crossings on the outside
part of the FS, while the FS break remains sharp in the direction of $\Gamma$. 
This effect of smearings on the inner and outer sides of the FS is attenuated if $\xi$ decreases with
increasing $q$, but as was mentioned, effects from the estimated variation of $\xi$ are small  
in comparison to variations of $q$. 

\begin{figure}
\includegraphics[height=5.5cm,width=7.0cm]{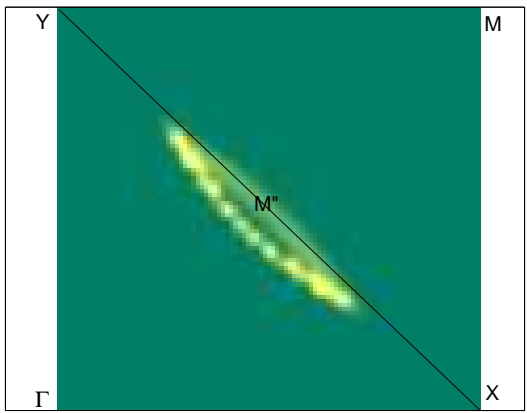}
\caption{(Color online) An intensity plot (the occurence of having the bands within 2 mRy of $E_F$
for each $k_x,k_y$-bin) for the folded FS of LCO with $E_F$ shifted
to be close to optimal doping. The exchange is 10-15 mRy with $q$ from 0.04 to 0.012.
}
\label{fig5}
\end{figure}
\begin{figure}
\includegraphics[height=5.5cm,width=7.0cm]{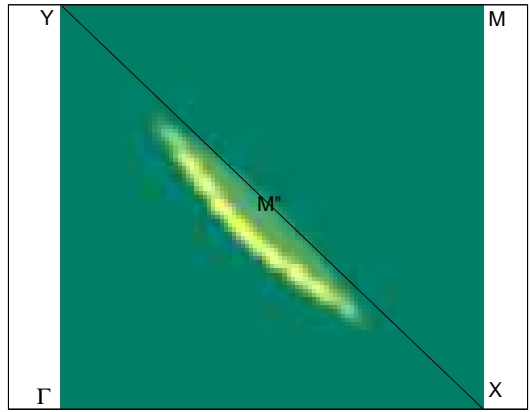}
\caption{(Color online) As in Fig. \ref{fig5}, but for 3 mRy lower $E_F$, with
$\xi$ 8-12 mRy and  $q$ 0.08-0.03.  
}
\label{fig6}
\end{figure}

An average of the FS signal is made from 
a set of about 20 equally spaced q-vectors.  
The bands in Fig. \ref{fig1} are folded into 
the zones of increasing $q$. The range of $q$- and $\xi$-values is a factor of 3 and
$\pm$20 percent, respectively, as was concluded above. Fig. \ref{fig5} shows an example with the $\xi=
12.5 \pm 2.5$ mRy and $q$ in the range [0.012-0.04]. At these conditions  a
trace of the outermost FS remains near $M"$, with a minimum of the intensity closer to $\Gamma$,
between the inner and outer FS-parts, similar to what was observed by Meng {\it et al}
for underdoped Bi-2201 \cite{meng}. The main inner peak is even
more dominant when spin fluctuations are weak or absent. 
This is because at weaker spin fluctuations there are larger $q$ and larger relative spread $\Delta q$
which contribute to a diffuse outer part of the FS. 
(Oppositely, $q \rightarrow 0$ if only strong fluctuations
are present so that also $\Delta q \rightarrow 0$.)
This situation is approached in Fig. \ref{fig6}. 
Smaller $\xi$ implies larger $\vec{q}$, and the
widths of q-values make stronger smearing of the outer branch, and no
minumum is seen between the two FS branches in Fig. \ref{fig6}. 
The main inner peak appears sligthly closer toward $\Gamma$ in Fig. \ref{fig6} than
in Fig. \ref{fig5}, because of the general expansion of the FS barrel for increased
hole doping. The central region for maximum intensity from the outer FS happens to be near
$M"$ in Figs. \ref{fig5} and \ref{fig6}, but there is no fundamental reason for this.  
Remnants of the outer FS are identified near or slightly inside the $M"$ positions
in ARPES for two underdoped compositions
of Bi2201, and a
very broad and unstructured distribution has been observed for optimal doping \cite{meng}. 
Finally, if $\xi \rightarrow 0$ then the non-magnetic (unfolded) BZ is appropriate with only
the inner FS branch.

Spin fluctuations are strong at low doping before static AFM order occurs in undoped 
insulating cuprates. On the contrary,
at high doping it is believed that fluctuations and $\xi$
decrease to a point when a representation of the FS within the NM BZ becomes appropriate. 
Of course,
only the inner branch of the FS exists for the NM case. 
The inner FS part from
weak fluctuations will overlap with the single FS of the NM FS, while the outer part from weak
fluctuations will be faint and broad. Therefore, the observation in ref. \cite{meng}
for the highest doping might be a case toward weakened striped fluctuations with a dominant signal from the 
NM FS. 

The independent work of Harrison {\it et al} \cite{harr} relies on a model of 
Lee {\it et al} \cite{lee} for a Lorenzian
probability distribution for the diffusion, where wave vectors and a spin correlation
length $\ell$ are parameters. By deviating the wave vector from the zone limit they
obtain a displacement of the FS orbit, as in the present work. The extreme limits of
their FS smearing is introduced
by varying $\ell$ from 100 to 5 \AA. The similarity of the two approaches 
is evident for the q-dependence, while in the present work smearing is introduced
by the step-like range of q-vectors. An estimate of the correlation length is not directly 
comparable with
$\ell$ in ref \cite{harr}, because of the different distributions and cuf-offs in real 
and reciprocal space, but if $\ell \approx L \approx \pi / \Delta q$ and $\Delta q$ are 
0.028 and 0.05 in Figs. \ref{fig5} and \ref{fig6}, we can deduce correlation lengths of
roughly 120-60 \AA~ as an order of magnitude of the lengths in these two cases.   
Harrison {\it et al} derived damping properties
for quantum oscillation from these parametrized calculations for different correlation
lengths. A corresponding analysis will not be undertaken here, but because of
the similarities of the two approaches it can be assumed that the damping properties
of quantum oscillation will be close to the results of ref \cite{harr} for rather large $\ell$.

\begin{figure}
\includegraphics[height=5.5cm,width=7.0cm]{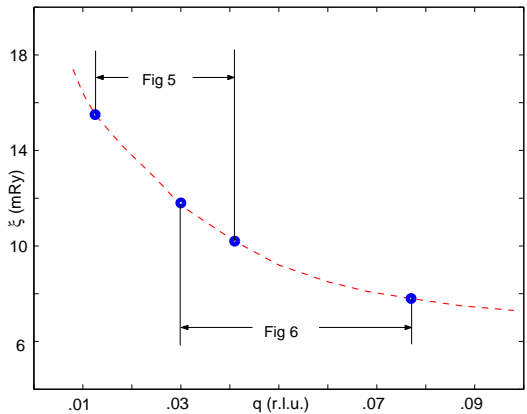}
\caption{(Color online) Approximate ranges of the exchange splitting, $\xi$, as function of modulation
vector, $q$, used for Figures \ref{fig5} and \ref{fig6}.  
}
\label{fig7}
\end{figure}

The present model includes a uniform distribution of fluctuations within the ($q,\xi$)-interval, with no
contributions outside. For instance, Fig. \ref{fig6} is based on 16 equally spaced ($q,\xi$)-values
starting from (0.03,12) and ending at (0.08,8) (in units of $r.l.u$ and $mRy$), see Fig. \ref{fig7}.
Real fluctuations from 
O-modes with $\hbar \omega$ around
50 meV contribute mostly to the low-$q$, high-$\xi$ part of the ranges (to the left in Fig. \ref{fig7}) \cite{tj9}, 
and such particular ($q,\xi$)-values would show increased intensity in neutron data 
\cite{vig,tran1,mat,hin}. 
Such variations could modify the details of the appearance of ARPES intensities \cite{qo}, but 
the extreme case with only one ($q,\xi$)-value would make the breaks of the inner 
and outer parts of the FS
equally sharp. 
However, possible effects of limited correlation lengths remain. The $(q,\xi)$-range in Fig. \ref{fig7} will,
according to the model for spin-phonon coupling, move
more to the right and be wider if the hole-doping increases  \cite{tj7,tj9}.

Finally, a comment about the charge character of the small $M"$ centered orbit. From Fig. \ref{fig2} it
appears to be hole like when the pocket is very small. 
However, the charge character depends on the curvature of the band $\epsilon_k$,
i.e. $d^2\epsilon_k/dk^2$ \cite{akp}. As is seen in Fig. \ref{fig2} between
$X$ and $M"$ there is an increasingly positive
second derivative of the band when $E_F$ gets lower from hole doping, which is
a signature of electronic character. The band is rather flat
in this portion of the FS, so it contributes significantly to the DOS and it may explain the 
unexpected electron-like character of the band in Hall measurements \cite{leBoeuf}. If so,
from the dispersion of the band in Fig. \ref{fig2} it can be expected that the electron character gets weaker 
for less hole doping, i.e. when the area of the orbit becomes very small. An alternative explanation
is that rests of the X-pocket are contributing even at intermediate doping \cite{leBoeuf}.

\section{Conclusion}

 The FS of a hole doped AFM cuprate has a banana shaped hole pocket centered slightly inside the
$M"$-point.  In addition to effects from the finite correlation length of
spin fluctuations \cite{harr}, it is shown here that stripe-like modulations of spin fluctuations 
with multiple $q$-values make the outer part
of this FS diffuse, while the inner part remains sharp. 
Only the latter part is visible for weak or nonexistent spin fluctuations. 
A distribution of different
exchange splittings is important for asymmetric broadening through the connection to
variations of $q$. 
Both lifetime broadening and
static fluctuations of different strengths within domains lead to asymmetric broadening. 
A natural connection between results of ARPES \cite{meng,nor,dama} and neutron scattering
\cite{vig,tran1,mat,hin} is suggested.
Furthermore, spin-phonon coupling has been shown
to be compatible with the ($q,\omega$)-dependence of spin excitations, 
and it might be an important ingredient for the 
mechanisms behind pseudogaps and superconductivity \cite{tj9}. 
A concordance between ARPES intensities and spin excitations seen by
neutrons indicates that spin-phonon
coupling is at work in the cuprates.


\begin{thebibliography}{10}
\bibitem{qo} Grouping of q-vectors around particular values will be favorable for quantum oscillations
because of smaller effective $\Delta q$.
\end{thebibliography}
\end{document}